\documentclass[preprint,pra]{revtex4}
\begin{document}
\input{epsf}


\title{ Focusing Vacuum Fluctuations II}

\author{L.H. Ford} 
 \email[Email: ]{ford@cosmos.phy.tufts.edu} 
 \affiliation{Institute of Cosmology, Department of Physics and Astronomy// 
         Tufts University, Medford, MA 02155}

\author{N.F. Svaiter } 
 \email[Email: ]{svaiter@lns.mit.edu } 
 \altaffiliation[\protect\\ Permanent address: ]{ Centro Brasileiro de 
Pesquisas Fisicas CBPF, Rua Dr. Xavier Sigaud 150,
Rio de Janeiro, RJ, 22290 180, Brazil } 
 \affiliation{Center for Theoretical Physics,
Laboratory for Nuclear Science and Department of Physics,
Massachusetts Institute of Technology, Cambridge, MA 02139}

\begin{abstract}
The quantization of the scalar and electromagnetic fields in the presence of
a parabolic mirror is further developed in the context of a geometric
optics approximation. We extend results in a previous paper to more general
geometries, and also correct an error in one section of that paper.
 We calculate the mean squared scalar and electric
fields near the focal line of a parabolic cylindrical mirror. These quantities
are found to grow as inverse powers of the distance from the focus. 
We give a combination of analytic and numerical results for the mean
squared fields. In particular, we find that the mean squared electric
field can be either negative or positive, depending upon the choice of
parameters. The case of a negative mean squared electric field corresponds
to a repulsive Van der Waals force on an atom near the focus, and to a
region of negative energy density. Similarly, a positive value corresponds
to an attractive force and a possibility of atom trapping in the vicinity 
of the focus.
\end{abstract}
 

\maketitle 

\baselineskip=14pt

\section{Introduction}
\label{sec:intro}

In a previous paper \cite{FS00}, henceforth I, we developed a geometric 
optics approach to the quantization of scalar and electromagnetic fields
near the focus of a parabolic mirror. We found that there can be enhanced
fluctuations near the focus in the sense that mean squared field
quantities scale as an inverse power of the distance from the focus,
rather than an inverse power of the distance from the mirror. These
enhanced fluctuations were found to arise from an interference term
between different reflected rays. In the present paper we extend 
the previous treatment. In I, only points on the symmetry axis were considered.
Here we are able to treat points in an arbitrary direction from the
focal line of a parabolic cylinder. We give more detailed
numerical results which provide a fuller picture of the phenomenon
of focusing of vacuum fluctuations. We also correct some erroneous results
in Sect.~V of I. 

The outline of the present paper is as follows: In Sect.~\ref{sec:Formalism},
we review and extend some of the formalism used to calculate the
mean squared scalar and electric fields, $\langle \varphi^2 \rangle$ and
$\langle {\bf E}^2 \rangle$, near the focus. In Sect.~\ref{sec:para}
we derive some geometric expressions which are needed to study fluctuations
at points off of the symmetry axis. In Sect.~\ref{sec:sing}, the
evaluation of integrals with singular integrands is revisited. Two
different, but equivalent, approaches are discussed. A particular case
where the integrals can be performed analytically is examined in 
Sect.~\ref{sec:exact}. More generally, it is necessary to calculate
$\langle \varphi^2 \rangle$ and $\langle {\bf E}^2 \rangle$  numerically.
A procedure for doing so is outlined in Sect.~\ref{sec:num}. Some detailed 
numerical results are also presented there. The limits of validity
of our model and results will be examined in Sect.~\ref{sec:limits}.
This discussion will draw on some results on diffraction obtained in the
Appendix. The experimental testability of our conclusions will be discussed 
in Sect.~\ref{sec:obs}. Finally, the results of the paper will be summarized 
in Sect.~\ref{sec:final}.

 Units in which $\hbar =c = 1$ will be used throughout this paper.
Electromagnetic quantities will be in Lorentz-Heaviside units. 

\section{Basic Formalism}
\label{sec:Formalism}

Here we will briefly review the geometric optics approach developed
in I. The basic assumption is that we may use a ray tracing method to
determine the functional form of the high frequency modes, which will
in turn give the dominant contribution to the expectation values of 
squared field operators. We start with a basis of plane wave modes. The
incident wave, for a scalar field, may be taken to be
\begin{equation}
f_{\bf k} = \frac{1}{\sqrt{2 \omega V}} \, 
{\rm e}^{i({\bf k}\cdot{\bf x}-\omega t)} \,,             \label{eq:plane}
\end{equation} 
with box normalization in a volume $V$. In the presence of a boundary,
this is replaced by the sum of incident and reflected waves,
\begin{equation}
F_{\bf k} = f_{\bf k} + \sum_i  f^{(i)}_{\bf k} \, , \label{eq:mode}
\end{equation}
where the $f^{(i)}_{\bf k}$ are the various reflected waves. One could
also in principle adopt a wavepacket basis, in which $F_{\bf k}$ is 
replaced by a localized wavepacket. Because the time evolution preserves the 
Klein-Gordon norm, if the various modes are orthonormal in the past,
they will remain so after reflection from the mirror. Thus we can view
Eq.~(\ref{eq:mode}) as the limit of a set of orthonormal wavepacket modes
in which the modes become sharply peaked in frequency and hence delocalized.

It was shown in I that the renormalized expectation value of the squared
scalar field is given by a sum of interference terms:
\begin{equation}
\langle \varphi^2 \rangle = 
\sum_{\bf k} \left[\sum_i (f^*_{\bf k} f^{(i)}_{\bf k}
+f_{\bf k} {f^{(i)}}^*_{\bf k} ) 
+ \sum_{i \not= j} f^{(i)}_{\bf k} {f^{(j)}}^*_{\bf k} \right] \,.
\label{eq:phisqren}
\end{equation}
This renormalized expectation value is defined as a difference in the mean 
value of $\varphi^2$ with and without the mirror, and hence will vanish at
large distances from the mirror. The various interference terms in the 
above expression yield contributions to $\langle \varphi^2 \rangle$
which are proportional to the inverse square of the appropriate path 
difference. Thus in the vicinity of the focus, the interference terms
between different reflected rays will dominate over that between the
incident and a reflected ray. In the present paper, we will consider cases
with no more than two reflected rays, and write
\begin{equation}
\langle \varphi^2 \rangle \approx 
2\, {\rm Re} \sum_{\bf k} f^{(1)}_{\bf k} {f^{(2)}}^*_{\bf k}  \,.
\label{eq:phisqren2}
\end{equation}

In the case of a parabolic cylinder, this may be expressed as
\begin{equation}
\langle \varphi^2 \rangle = -\frac{1}{3 \pi^2} \int 
                          \frac{d \theta'}{(\Delta \ell)^2} \,.
\label{eq:phisqren3}
\end{equation}
Here $\Delta \ell$ is the path difference for the two reflected rays,
and the integration is over the reflection angle of one of the rays.
The range is chosen so that each pair of reflected rays is counted once.
The corresponding expression for the mean squared electric field near
the focus of a parabolic cylinder is found in I to be
\begin{equation}
\langle {\bf E}^2 \rangle =  \frac{8}{5 \pi^3} \int 
                          \frac{d \theta'}{(\Delta \ell)^4}\, .  
\label{eq:Esqren}
\end{equation}

\section{Optics of Parabolic Mirrors}
\label{sec:para}

\begin{figure}
\begin{center}
\leavevmode\epsfysize=8cm\epsffile{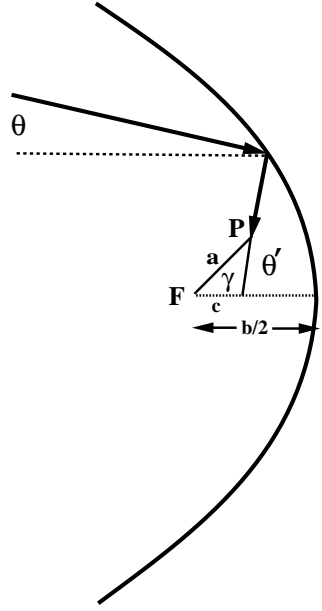}
\end{center}
\caption{An incident ray at an angle $\theta$ reflects off of a parabolic
mirror at an angle $\theta'$ and arrives at the point of interest $P$.
This point is a distance $a$ from the focus $F$ in a direction at an angle
$\gamma$ with respect to the symmetry axis. Throughout this paper, we
assume that $a \ll b$, where $\frac{1}{2} b$ is the distance from the focus
to the mirror. }
\label{fig:para}
\end{figure}

In this section, we wish to generalize some of the results of I concerning
the incident and reflected rays in the presence of a parabolic mirror. Consider
the geometry illustrated in Fig.~\ref{fig:para}. An incident ray at an
angle of $\theta$ to the symmetry axis is reflected from the point $(x_i,y_i)$,
and then reaches the point $P$ at an angle of $\theta'$. We first need to find 
the relation between $\theta$ and $\theta'$. Note that the reflected ray
crosses the symmetry axis at a distance $c$ from the focus. It was shown
in I that
\begin{equation}
\theta = \frac{c\, \sin^3 \theta' }{b (1 - \cos \theta' )} \,.
                                                     \label{eq:theta0}
\end{equation}
However, from the law of sines, we have that 
\begin{equation}
c\, \sin \theta' = a\,\sin (\theta'-\gamma) \,. 
\end{equation}
Hence we now have that 
\begin{equation}
\theta = \frac{a}{b} \, f(\theta') \,,
\end{equation}
where 
\begin{equation}
f(\theta') = \frac{\sin^3 \theta' \, \sin (\theta'-\gamma)}
                                  {(1 - \cos \theta' )} \,. \label{eq:theta}
\end{equation}  

\begin{figure}
\begin{center}
\leavevmode\epsfysize=8cm\epsffile{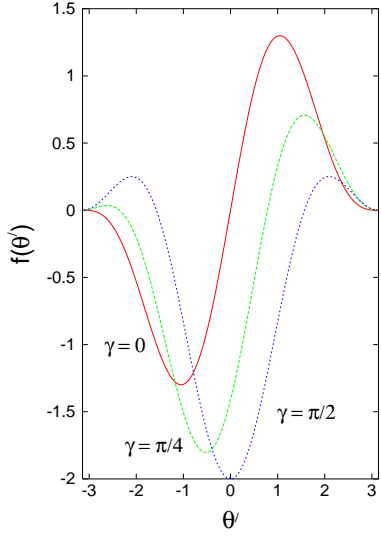}
\end{center}
\caption{The function $f(\theta')$ is plotted for various values of $\gamma$.
This function relates the angle of the incident ray, $\theta$, to the angle
of the reflected ray, $\theta'$, through the relation 
$\theta = (a/b)\, f(\theta')$. }
\label{fig:f}
\end{figure}

There will be multiply reflected rays whenever different values of $\theta'$
are associated with the same value of $f$.
The function $f(\theta')$ is plotted in Fig.~\ref{fig:f} for various values of 
$\gamma$. We can see from these plots that in general there can be up to
four reflected angles $\theta'$ for a given incident angle $\theta$.
However, if the mirror size $\theta_0$ is restricted to be less than $2 \pi/3$,
then there will never be more than two values of $\theta'$ for a given $\theta$.
Throughout this paper, we will assume $\theta_0 < 2 \pi/3$, and hence have
at most two reflected rays for a given incident ray. The two reflected rays will 
occur at $\theta'=\alpha$ and $\theta'=\beta$, where
\begin{equation}
f(\alpha) = f(\beta) \, .
\end{equation}

\begin{figure}
\begin{center}
\leavevmode\epsfysize=8cm\epsffile{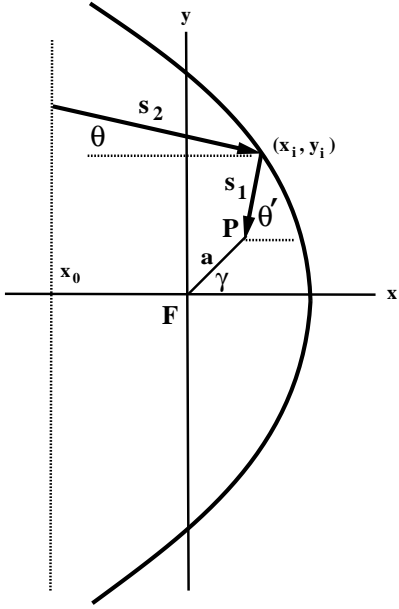}
\end{center}
\caption{The distance travelled by a ray which reaches the point $P$ is 
illustrated. Here the line $x = x_0$ is an arbitrarily chosen vertical line
somewhere to the left of the point of reflection. The incident ray travels
a distance $s_2$ from this line to the point of reflection. After reflection
it travels a distance $s_1$ before arriving at the point $P$. This point is 
a distance $a$ from the focus $F$ in a direction given by the angle $\gamma$. }
\label{fig:para2}
\end{figure}

Our next task is to calculate the difference in path length, $\Delta \ell$,
 for these two reflected rays. Again, this is a generalization of a calculation
given in I. Consider the situation illustrated in Fig.~\ref{fig:para2},
where a reflected ray with angle $\theta'$ reaches the point $P$ after 
reflecting from the point $(x_i,y_i)$ on the mirror. Let $s_1$ be the
distance traveled after reflection, and $s_2$ be the distance traveled
between when the ray crosses the line $x=x_0$ and when it reaches the mirror.
Note that
\begin{equation}
s_2 = (x_i - x_0) \sec \theta \approx x_i - x_0 + O\left(\frac{a^2}{b^2}\right) \,.
\end{equation}
Even if $x_i < 0$, we can choose $x_0$ to be such that $s_2 > 0$.
Next note that
\begin{equation}
s_1 = (y_i -a \sin \gamma) \csc \theta' \,.
\end{equation}
The reflection point $(x_i,y_i)$ is the intersection of the line
\begin{equation}
y = \tan \theta' (x -a\cos \gamma) + a \sin \gamma
\end{equation}
with the parabola, given by
\begin{equation}
x = \frac{b^2-y^2}{b} \,.
\end{equation}
These relations lead to
\begin{equation}
y_i = b \cot \theta' \left[ \sec \theta' \sqrt{1 -2\left(\frac{a}{b}\right)
\sin^2\theta' (\cos\gamma -\sin\gamma \cot\theta')} -1 \right] \,.
\end{equation}
We may now expand this expression to first order in $a/b$ and combine it with
our previous expressions to show that, to first order, 
\begin{equation}
\ell = s_1 + s_2 = b -x_0 -a(\cos\gamma \cos\theta' - \sin\gamma \sin\theta')\,.
\end{equation}
Thus the magnitude of the path length difference for $\alpha$ and
$\beta$, two different values of $\theta'$, is
\begin{equation}
\Delta \ell = a\,|\cos\gamma (\cos\alpha -\cos\beta)
                  + \sin\gamma (\sin\alpha -\sin\beta)| \,.
\end{equation}
In the limit that $\gamma =0$, we obtain the result for $\Delta \ell$ used
in I.

\section{Evaluation of Singular Integrals}
\label{sec:sing}

We can express Eq.~(\ref{eq:phisqren3}) as
\begin{equation}
\langle \varphi^2 \rangle = -\frac{1}{6 \pi^2\, a^2} \int 
                          \frac{d \alpha}{h^2} \, ,
\label{eq:phisqren4}
\end{equation}
and Eq.~(\ref{eq:Esqren}) as
\begin{equation}
\langle {\bf E}^2 \rangle =  \frac{4}{5 \pi^3 \, a^4} \int 
                          \frac{d \alpha}{h^4}\, .  
\label{eq:Esqren2}
\end{equation}
Here
\begin{equation}
h = h(\alpha,\beta) = |\cos\gamma (\cos\alpha -\cos\beta)
 + \sin\gamma (\sin\alpha -\sin\beta)| \, ,  \label{eq:h}
\end{equation}
and the integrations run over the full range for which there are two
reflected rays, thus counting each pair twice.

The integrands of these integrals contain singularities within the range
of integration. These singularities occur at a critical angle, $\theta_c$,
at which $h$ vanishes linearly. Thus the $\langle \varphi^2 \rangle$
integral contains a $(\alpha -\theta_c)^{-2}$ singularity, and the
$\langle {\bf E}^2 \rangle$ contain a $(\alpha -\theta_c)^{-4}$ singularity.
These singularities are presumably artifacts of assuming perfect mirrors with
sharp boundaries as will be discussed in Sect.~\ref{sec:limits}.
 The singularity occurs when both $\alpha$ and $\beta$
are approaching $\theta_c$ from opposite directions. The critical angles
$\theta_c$ are just the extrema of the function $f(\theta')$.  

Integrals with such singular integrands can be defined by a generalization
of the principal value prescription. This generalization involves an integration 
by parts to recast the original integral as one containing a less singular
integrand, plus surface terms. For example, we may use
\begin{equation}
\frac{1}{x^2} = - \frac{1}{2} \frac{d^2}{d x^2} \ln x^2   \label{eq:inv_x2}
\end{equation}
to write
\begin{equation}
\int_a^b dx\, \frac{f(x)}{x^2} = - \frac{1}{2}\int_a^b dx\,f''(x) \, \ln x^2\,
\quad +\left[-\frac{f(x)}{x} +\frac{1}{2} f'(x)\, \ln x^2 \right]_a^b  \,.
                                     \label{eq:x2}
\end{equation}
Similarly,
\begin{equation}
\frac{1}{x^4}= - \frac{1}{12}\, \frac{d^4}{d x^4} \ln x^2 \, ,  \label{eq:inv_x4}
\end{equation}
leads to
\begin{equation}
\int_a^b dx\, \frac{f(x)}{x^4} = -\left[\frac{f(x)}{3x^3}+\frac{f'(x)}{6x^2}
+\frac{f''(x)}{6x} -\frac{1}{12} f'''(x)\,\ln x^2 \right]_a^b
-\frac{1}{12}\int_a^b dx\,\ln x^2\,\frac{d^4 f(x)}{d x^4} \,. \label{eq:x4}
\end{equation}
If $b>0$ and $a<0$, there were singularities in the original integrals
which are replaced by integrable, logarithmic singularities. In all cases,
the surface terms are evaluated away from the singularity and are hence
finite. Note that we could have added additional polynomial terms on the
righthand sides of both Eqs.~(\ref{eq:inv_x2}) and (\ref{eq:inv_x4}). 
For example,
we could replace $\ln x^2$ by $\ln x^2 + c_1 x + c_0$ in Eq.~(\ref{eq:inv_x2}). 
However, the arbitrary constants $c_0$ and $c_1$ will cancel out between
the two terms on the righthand side of Eq.~(\ref{eq:x2}) and hence can be 
ignored. A similar cancellation occurs if we add additional terms into
Eq.~(\ref{eq:inv_x4}).

If $f$ and a sufficient number of its derivatives vanish at the endpoints,
then the surface terms vanish. In I, it was incorrectly argued that the 
surface terms which arise in the present problem can be ignored. It is true
that if the reflectivity of the mirror falls smoothly to zero, the surface
terms vanish. However, in this case it would be necessary to integrate
explicitly the function which is falling smoothly to zero. If the reflectivity
falls very rapidly at the edge of the mirror, its derivatives will be large,
and effectively reproduce the surface terms. Thus, the explicit results
given in Sect.~5 of I are incorrect. In Sects.~\ref{sec:exact} and
\ref{sec:num} of the present paper, we will give new results which replace 
and generalize the older results.

There is an alternative way to implement the above integration by parts
prescription. This is simply to evaluate the integral as an indefinite 
integral, and evaluate the result at the endpoints, ignoring the singularity
within the integration range. Even if it is not possible to find a closed
form expression for the indefinite integral, one can expand the integrand
in a Laurent series about $x=0$ and integrate term by term, using the relation
\begin{equation}
 \int_a^b \frac{1}{x^n}\, dx = \frac{1}{n+1} \left(\frac{1}{a^{n+1}}
    - \frac{1}{b^{n+1}} \right) \,.    \label{eq:directint}
\end{equation}
One can verify this relation for the case $a < 0 < b$ using the above 
integration by parts method.

The integrands in Eqs.~(\ref{eq:phisqren4}) and (\ref{eq:Esqren2}) 
arise from the integrals
\begin{equation}
\frac{1}{x^2} = - \int_0^\infty d\omega \, \omega \, \cos\omega x
\end{equation}
and
\begin{equation}
\frac{1}{x^4} = \frac{1}{6} \int_0^\infty d\omega^3 \, \omega \, \cos\omega x \,,
\end{equation}
respectively. The integrals are understood to be performed with the aid of a 
convergence factor, such as ${\rm e}^{-\lambda \omega}$, with the limit
$\lambda \rightarrow 0 $ taken after integration. The singularities
at $x=0$ are due to the contributions of arbitrarily large values of
$\omega$. If there were to be a physical cutoff at high frequencies,
then these singulariries would disappear. However, the integrals containing
the singular integrands are independent of the cutoff in the limit that
it occurs at sufficiently high frequencies. As an example, consider the 
function 
\begin{equation}
g_2(x,\lambda) = \int_0^\infty d\omega \, \omega \, 
{\rm e}^{-\lambda \omega}\, \cos\omega x = 
\frac{\lambda^2 -x^2}{(\lambda^2 +x^2)^2} \, .
\end{equation}
As $\lambda \rightarrow 0$, $g_2(x,\lambda) \rightarrow -1/x^2$, but
for $\lambda \not= 0$, $g_2(x,\lambda)$ is finite for all $x$. Furthermore,
if we integrate $g_2(x,\lambda)$ on $x$ through $x=0$, and then take
the $\lambda \rightarrow 0$ limit, the result will be finite and the same
as that obtained by the above formal procedures:
\begin{equation}
\int_{-x_0}^{x_0} g_2(x,\lambda) d x = \frac{2 x_0}{\lambda^2 +x_0^2}
\rightarrow \frac{2}{x_0} \, .
\end{equation}
Similarly, if we define
\begin{equation}
g_4(x,\lambda) = \int_0^\infty d\omega \, \omega^3 \, 
{\rm e}^{-\lambda \omega}\, \cos\omega x =  
\frac{6(x^2 -2\lambda x -\lambda^2)(x^2 +2\lambda x -\lambda^2)}
{(\lambda^2 +x^2)^4} \, ,
\end{equation}
then $g_4(x,\lambda) \rightarrow 6/x^4$ as $\lambda \rightarrow 0$.
However,
\begin{equation}
\int_{-x_0}^{x_0} g_4(x,\lambda) d x = 
-\frac{4 x_0 (x_0^2-3\lambda^2)}{\lambda^2 +x_0^3} 
\rightarrow -\frac{4}{x_0^3} \, .
\end{equation}
Thus we can understand the physical meaning of the formal integration
procedures discussed above as follows: they provide a shortcut method
to obtain the results one would obtain by inserting a frequency cutoff
in the $\omega$ integrals, and then removing this cutoff after integration 
over $x$.

One peculiar feature of our results will be that both 
$\langle \varphi^2 \rangle$ and 
$\langle {\bf E}^2 \rangle$ diverge for particular values of
$\gamma$. This arises when $\theta_c$, the extremum of $f(\theta')$,
approaches the edge of the mirror, $\theta'=\pm \theta_0$. The mathematical
reason for the divergence is that one limit of integration is approaching
a point at which $h = 0$ and the integrand is singular. This corresponds
to letting either $a \rightarrow 0$ with $b$ fixed, or $b \rightarrow 0$ 
with $a$ fixed, in Eq.~(\ref{eq:directint}). The physical origin of
this singular behavior is that we have made two unrealistic assumptions.
The mirror is assumed to be perfectly reflecting at all frequencies
and to have sharp edges at $\theta'=\pm \theta_0$. This will be discussed 
further in Sect.~\ref{sec:limits}.

\section{Exact Results for $\gamma = \frac{\pi}{2}$}
\label{sec:exact}

In general, the integrals for $\langle \varphi^2 \rangle$ and 
$\langle {\bf E}^2 \rangle$, Eqs.~(\ref{eq:phisqren3}) and (\ref{eq:Esqren}),
respectively, can only be evaluated numerically. This is in part because the
second reflection angle $\beta$ is implicitly given as a function of the 
first angle, $\alpha$. There is one case in which this relation may be
written down in closed form. This is when $\gamma = \frac{\pi}{2}$, so the
function $f(\theta')$ is symmetrical about the origin. In this case, we have
\begin{equation}
\beta = - \alpha \,,
\end{equation}
and we can write
\begin{equation}
\langle \varphi^2 \rangle = -\frac{1}{24 \pi^3 \, a^2} 
\int_{-\theta_0}^{\theta_0} \frac{d \alpha}{\sin^2 \alpha}
= \frac{\cot \theta_0}{12 \pi^3 \, a^2} \,. \label{eq:phi_0}
\end{equation}
Similarly,
\begin{equation}
\langle {\bf E}^2 \rangle = \frac{1}{20 \pi^3 \, a^4} 
\int_{-\theta_0}^{\theta_0} \frac{d \alpha}{\sin^4 \alpha}
= - \frac{\cos \theta_0 (3 - 2 \cos^2 \theta_0)}
         {30 \pi^3\, a^4 \, \sin^3\theta_0} \, .
\end{equation}

\begin{figure}
\begin{center}
\leavevmode\epsfysize=8cm\epsffile{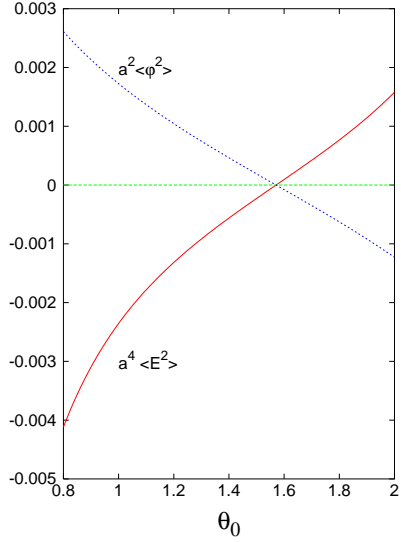}
\end{center}
\caption{The exact solutions for $\langle \varphi^2 \rangle$ and 
$\langle {\bf E}^2 \rangle$ for the case that $\gamma = \frac{\pi}{2}$
are plotted as functions of the mirror size. Note that 
$a^2 \langle \varphi^2 \rangle$ and $a^4 \langle {\bf E}^2 \rangle$ are
dimensionless in our units.  }
\label{fig:exact}
\end{figure}

These expressions are plotted in Fig.~\ref{fig:exact}.
Note that both $\langle \varphi^2 \rangle$ and $\langle {\bf E}^2 \rangle$
can have either sign:
\begin{eqnarray}
\langle \varphi^2 \rangle > 0 \quad {\rm and} 
\quad \langle {\bf E}^2 \rangle < 0
 \quad {\rm for} \quad  0 < \theta_0 < \frac{\pi}{2} \, , \nonumber \\
\langle \varphi^2 \rangle < 0 \quad {\rm and} 
\quad \langle {\bf E}^2 \rangle > 0
 \quad {\rm for} \quad \frac{\pi}{2}  < \theta_0 < \frac{2\pi}{3} \,. 
\end{eqnarray}
Both quantities seem to diverge in the limit that $\theta_0 \rightarrow 0$
\begin{eqnarray}
\langle \varphi^2 \rangle \sim \frac{1}{12 \pi^3 \, a^2\, \theta_0}\, ,
                                                      \nonumber \\
\langle {\bf E}^2 \rangle \sim - \frac{1}
         {30 \pi^3\, a^4 \, \theta_0^3} \, .
\end{eqnarray}
This divergence is a special case of the singularities arising when the
edge of the mirror approaches an extremum of $f(\theta')$.

We can go further and find the derivatives of both $\langle \varphi^2 \rangle$ 
and $\langle {\bf E}^2 \rangle$ with respect to $\gamma$, at 
$\gamma = \frac{\pi}{2}$. First, we need to find $\beta(\alpha)$ to
first order in $\gamma - \frac{\pi}{2}$. We may expand the relation
$f(\beta) = f(\alpha)$ to this order and show that
\begin{equation}
\beta = - \alpha - \frac{2 \sin^2 \alpha \,(\gamma - \frac{\pi}{2})}
{(\cos \alpha -1)(2 \cos \alpha +1)} + \cdots \,.
\end{equation}
We can next expand $1/h^2$ and $1/h^4$ to first order in 
$\gamma - \frac{\pi}{2}$. The first order corrections are odd functions of
$\alpha$ whose explicit forms we will not need. 

The crucial effect of varying $\gamma$ slightly away from $\pi/2$ is 
to change the range of integration. First consider the case where
$\gamma < \frac{\pi}{2}$. The integration range now becomes
\begin{equation}
-\theta_0 < \alpha < \theta_0 - \delta \, ,
\end{equation}
where
\begin{equation}
 \delta = \frac{2 \sin^2 \theta_0 \,(\frac{\pi}{2}-\gamma)}
{(1-\cos \theta_0 )(2 \cos \theta_0 +1)} \, .  \label{eq:delta}
\end{equation}
To first order in $\gamma - \frac{\pi}{2}$, we can write
\begin{eqnarray}
\langle \varphi^2 \rangle &=& -\frac{1}{6 \pi^3 \, a^2} 
\int_{-\theta_0}^{\theta_0 -\delta} \frac{d \alpha}{h^2} \nonumber \\
&=& -\frac{1}{6 \pi^3 \, a^2} \left[ 
\int_{-\theta_0}^{\theta_0} d \alpha \left(\frac{1}{h^2}\right)_0 
+ \int_{-\theta_0}^{\theta_0} d \alpha \left(\frac{1}{h^2}\right)_1
- \int_{\theta_0-\delta}^{\theta_0} d \alpha \left(\frac{1}{h^2}\right)_0
 - \cdots \right] \, , \label{eq:phi_expand}
\end{eqnarray}
where the subscripts on $1/h^2$ denote the order in an expansion in
powers of $\gamma - \frac{\pi}{2}$. The first term on the right hand side
of Eq.~(\ref{eq:phi_expand}) is just the zeroth order part given by
Eq.~(\ref{eq:phi_0}). The next term vanishes because $(1/h^2)_1$ is an 
odd function. The final term may be approximated using
\begin{equation}
\int_{\theta_0-\delta}^{\theta_0} d \alpha \left(\frac{1}{h^2}\right)_0
\approx \delta \, \left[ \left(\frac{1}{h^2}\right)_0 
                                      \right]_{\alpha=\theta_0} \, .
\end{equation}
This leads to
\begin{equation}
\langle \varphi^2 \rangle = \frac{1}{12 \pi^3 \, a^2}
\left[ \frac{1}{\tan \theta_0} - 
\frac{\frac{\pi}{2}-\gamma}{(1-\cos \theta_0)(2 \cos \theta_0 +1)} 
+ \cdots \right] \,, \qquad \gamma < \frac{\pi}{2} \, .
\end{equation}

We may now repeat this procedure for $\gamma > \frac{\pi}{2}$. In this case,
the range of integration is 
\begin{equation}
-\theta_0 - \delta < \alpha < \theta_0  \, ,
\end{equation}
and $\delta < 0$, so the range of integration has again decreased. This causes
the first order change in $\langle \varphi^2 \rangle$ to have the same magnitude
but the opposite sign from the previous case. Thus in all cases, we can
write
\begin{equation}
\langle \varphi^2 \rangle = \frac{1}{12 \pi^3 \, a^2}
\left[ \frac{1}{\tan \theta_0} - 
\frac{|\gamma - \frac{\pi}{2}|}{(1-\cos \theta_0)(2 \cos \theta_0 +1)} 
+ \cdots \right] \,.    \label{eq:phi_pi/2}
\end{equation}
A similar analysis may be applied to $\langle {\bf E}^2 \rangle$ with the
result
\begin{equation}
\langle {\bf E}^2 \rangle = \frac{4}{5 \pi^3 \, a^4}
\left[ - \frac{\cos \theta_0 (3 - 2 \cos^2 \theta_0)}{24 \, \sin^3\theta_0} 
- \frac{|\gamma - \frac{\pi}{2}|}
                     {8 \sin^2\theta_0 (1-\cos \theta_0)(2 \cos \theta_0 +1)} 
+ \cdots \right] \,.        \label{eq:E_pi/2}
\end{equation}

Thus both $\langle \varphi^2 \rangle$ and $\langle {\bf E}^2 \rangle$ have
cusps at $\gamma = \frac{\pi}{2}$. The nonanalytic behavior is due to the fact
that the range of integration decreases whenever $\gamma$ moves away from
$\frac{\pi}{2}$ in either direction. As will be discussed in 
Sect.~\ref{sec:limits}, the cusp is presumably an artifact of a sharp edge 
approximation, and should be smoothed out in a more exact treatment.

\section{Numerical Procedures and Results}
\label{sec:num}

Apart from the special cases discussed in the previous section, it is
necessary to evaluate $\langle \varphi^2 \rangle$ and 
$\langle {\bf E}^2 \rangle$
numerically. The first step is to find the second reflection angle $\beta$
as a function of the first reflection angle $\alpha$. This involves a
straightforward numerical solution of the equation
\begin{equation}
f(\beta) = f(\alpha) \,.   \label{eq:root}
\end{equation}
Because we assume a restriction on the angle size of the mirror, 
\begin{equation}
\theta_0 < \frac{2 \pi}{3} \, ,
\end{equation}
there will always be either one or no roots for $\beta$. Within the
geometric optics approximation that we use, integrands are assumed to
vanish in regions where there are no roots.  

As was discussed in Sect.~\ref{sec:sing}, there are at least two methods 
that may be used for explicit evaluation of the integrals which appear
in the geometric optics expressions for $\langle \varphi^2 \rangle$ and 
$\langle {\bf E}^2 \rangle$. The first is an integration by parts, which
replaces the singularity in Eqs.~(\ref{eq:phisqren4}) and (\ref{eq:Esqren2})
by a logarithmic singularity. Here we will outline how this may be done 
explicitly. Consider first  Eq.~(\ref{eq:phisqren4}), which may be expressed 
as
\begin{equation}
\langle \varphi^2 \rangle = -\frac{1}{6 \pi^2\, a^2} \int 
                          \frac{d \alpha}{[h(\alpha,\beta(\alpha))]^2} \, . 
\end{equation}
Suppose that we are interested in integrating over the range 
$\alpha_1 < \alpha < \alpha_2$
and that this range contains one point $\alpha = \alpha_0$ at which $h = 0$. We
can write
\begin{equation}
\langle \varphi^2 \rangle = \frac{1}{12 \pi^2\, a^2} 
\int_{\alpha_1}^{\alpha_2} d\alpha \;
\frac{(\alpha-\alpha_0)^2}{[h(\alpha,y(\alpha))]^2} \; 
\frac{d^2}{d\alpha^2} \log(\alpha-\alpha_0)^2 \,.
\end{equation}
The quantity $(\alpha-\alpha_0)^2/[h(\alpha,y(\alpha))]^2$ is finite at 
$\alpha = \alpha_0$. This expression
may now be integrated by parts, as in Eq.~(\ref{eq:x2}). The derivatives
of $h$ which arise are computed using Eq.~(\ref{eq:h}) and the relation
\begin{equation}
\frac{d\beta}{d\alpha} = \frac{f'(\alpha)}{f'(\beta)} \, .
\end{equation}
An analogous procedure can be applied to $\langle {\bf E}^2 \rangle$. 
The detailed expressions for $\langle \varphi^2 \rangle$ and 
$\langle {\bf E}^2 \rangle$
which result from this procedure are rather complicated, and will not written
down explicitly. 

The second method which may be employed is a variant of the direct integration
illustrated in Eq.~(\ref{eq:directint}). The actual integrands in
Eqs.~(\ref{eq:phisqren4}) and (\ref{eq:Esqren2}) are too complex to
integrate in closed form. However, we can use numerical integration
of $1/h^2$ and $1/h^4$ in regions away from zeros of $h$ and direct
integration of a series expansion in the neighborhood of a zero. The
first step in the generation of the series expansion is to expand $\beta(\alpha)$
around $\alpha = \alpha_0$, 
\begin{equation}
\beta = 2 \alpha_0 -\alpha +a_2(\alpha-\alpha_0)^2 
+a_3(\alpha-\alpha_0)^3 + \cdots \,.
\end{equation}
The coefficients are found by inserting this expansion into Eq.~(\ref{eq:root})
and then expanding both sides of the resulting expression in powers of
$\alpha-\alpha_0$. A few of the leading coefficients are
\begin{eqnarray}
a_2 &=& -\frac{f'''}{3 f''} \, , \nonumber \\
a_3 &=& -\frac{(f''')^2}{9 (f'')^2}\, , \nonumber \\
a_4 &=&  -\frac{9 (f'')^2 (f^{(5)})^2 -30 f'' f'''f^{(4)} +40(f''')^2}
             {540 (f'')^3} \,,
\end{eqnarray}
where the derivatives of $f$ are evaluated at $\alpha = \alpha_0$. 
The expansion
for $y(\alpha)$ is then used to generate analogous expansions for 
$1/h^2$ and $1/h^4$,
which are in turn integrated over the interval 
$\alpha_0 -\xi_0 < \alpha < \alpha_0 +\xi_0$
using Eq.~(\ref{eq:directint}). The result is combined with the direct numerical
integration outside of this interval. Here $\xi_0$ is an arbitrary small
positive number. One of the tests of the numerical procedure is the independence
of the results upon the choice of $\xi_0$.

We have developed numerical routines based upon both of the above procedures.
The plots which are
given below were created using a routine based upon the second method, with
an expansion of $y$ to sixth order in 
$\alpha-\alpha_0$. The particular values of $\xi_0$
were in the range $0.1 \leq \xi_0 \leq 0.3$. In this range, the routine is
relatively stable and insensitive to $\xi_0$. Smaller values of $\xi_0$
can lead to instability, as the contributions from $|x-x_0| > \xi_0$
and from $|x-x_0| < \xi_0$ are both large in magnitude and tending to cancel
each other.

\begin{figure}
\begin{center}
\leavevmode\epsfysize=8cm\epsffile{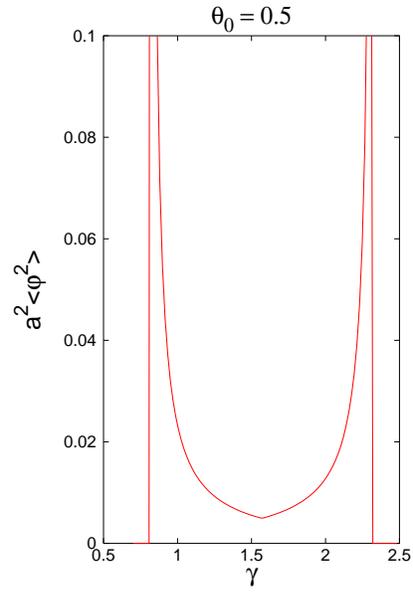}
\end{center}
\caption{$\langle \varphi^2 \rangle$ for $\theta_0 = 0.5$. }
\label{fig:para_p_50}
\end{figure}

\begin{figure}
\begin{center}
\leavevmode\epsfysize=8cm\epsffile{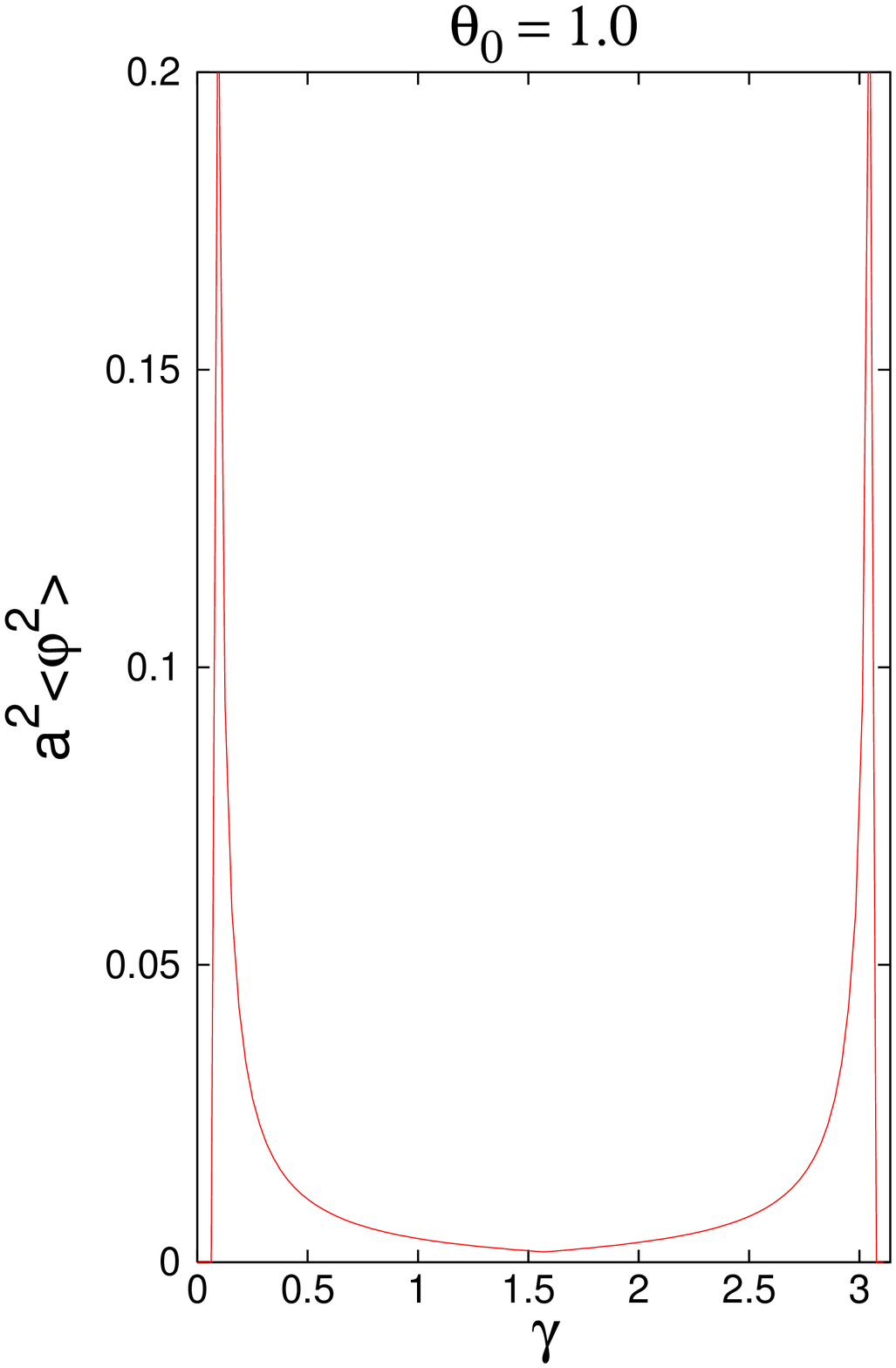}
\end{center}
\caption{$\langle \varphi^2 \rangle$ for $\theta_0 = 1.0$. }
\label{fig:para_p_100}
\end{figure}

\begin{figure}
\begin{center}
\leavevmode\epsfysize=8cm\epsffile{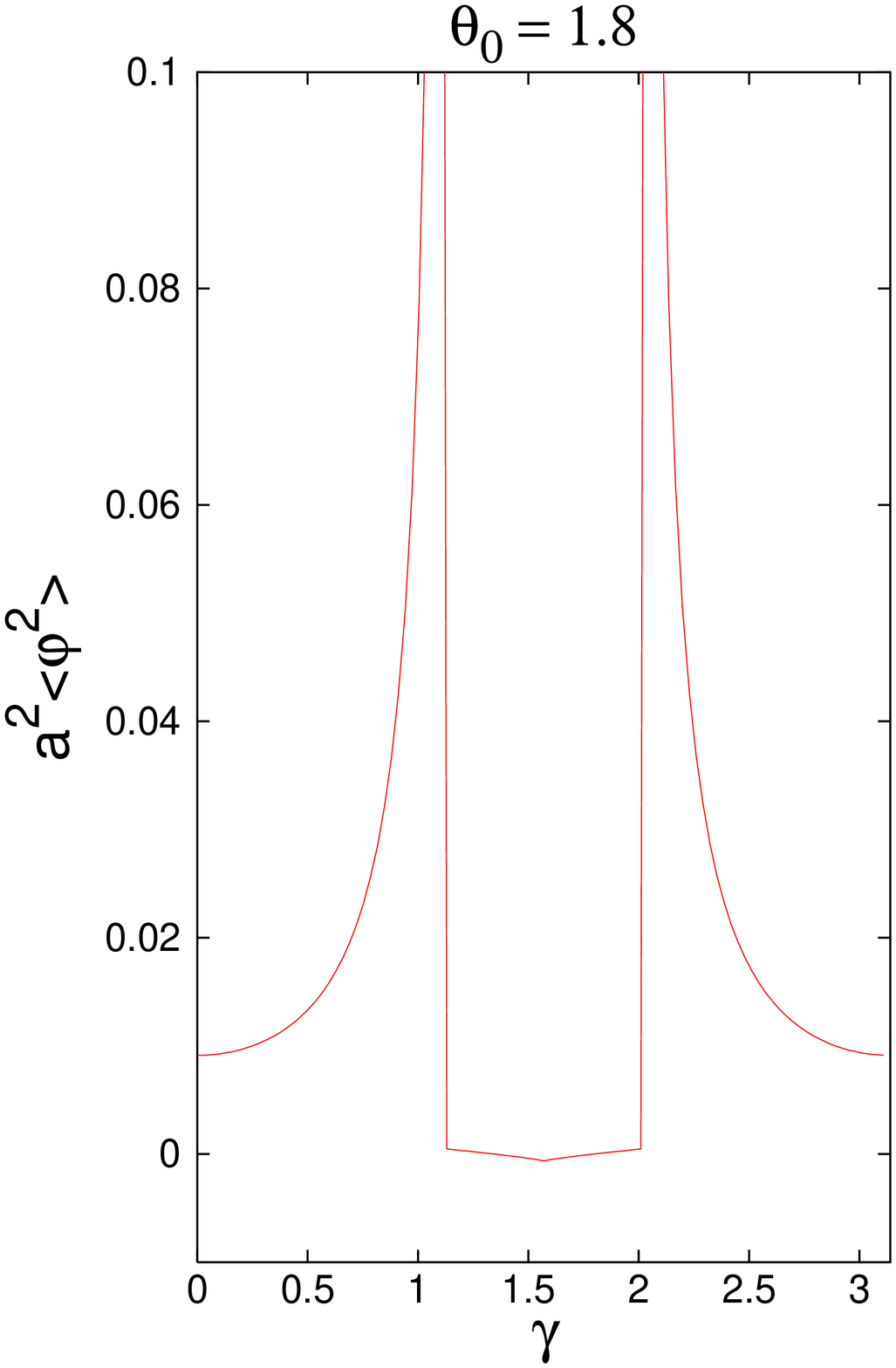 }
\end{center}
\caption{$\langle \varphi^2 \rangle$ for $\theta_0 = 1.8$. }
\label{fig:para_p_180}
\end{figure}

\begin{figure}
\begin{center}
\leavevmode\epsfysize=8cm\epsffile{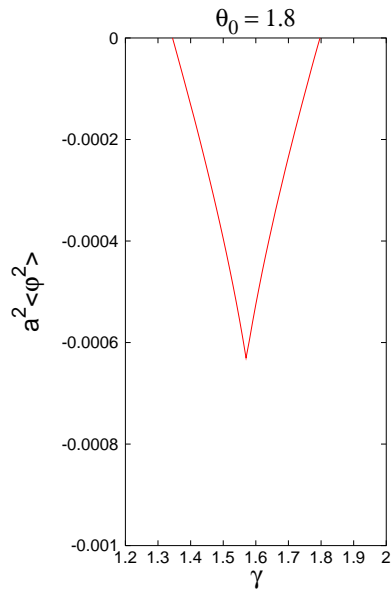 }
\end{center}
\caption{$\langle \varphi^2 \rangle$ for $\theta_0 = 1.8$, showing in detail 
the region where $\langle \varphi^2 \rangle <0$. }
\label{fig:para_p_180_2}
\end{figure}

\begin{figure}
\begin{center}
\leavevmode\epsfysize=8cm\epsffile{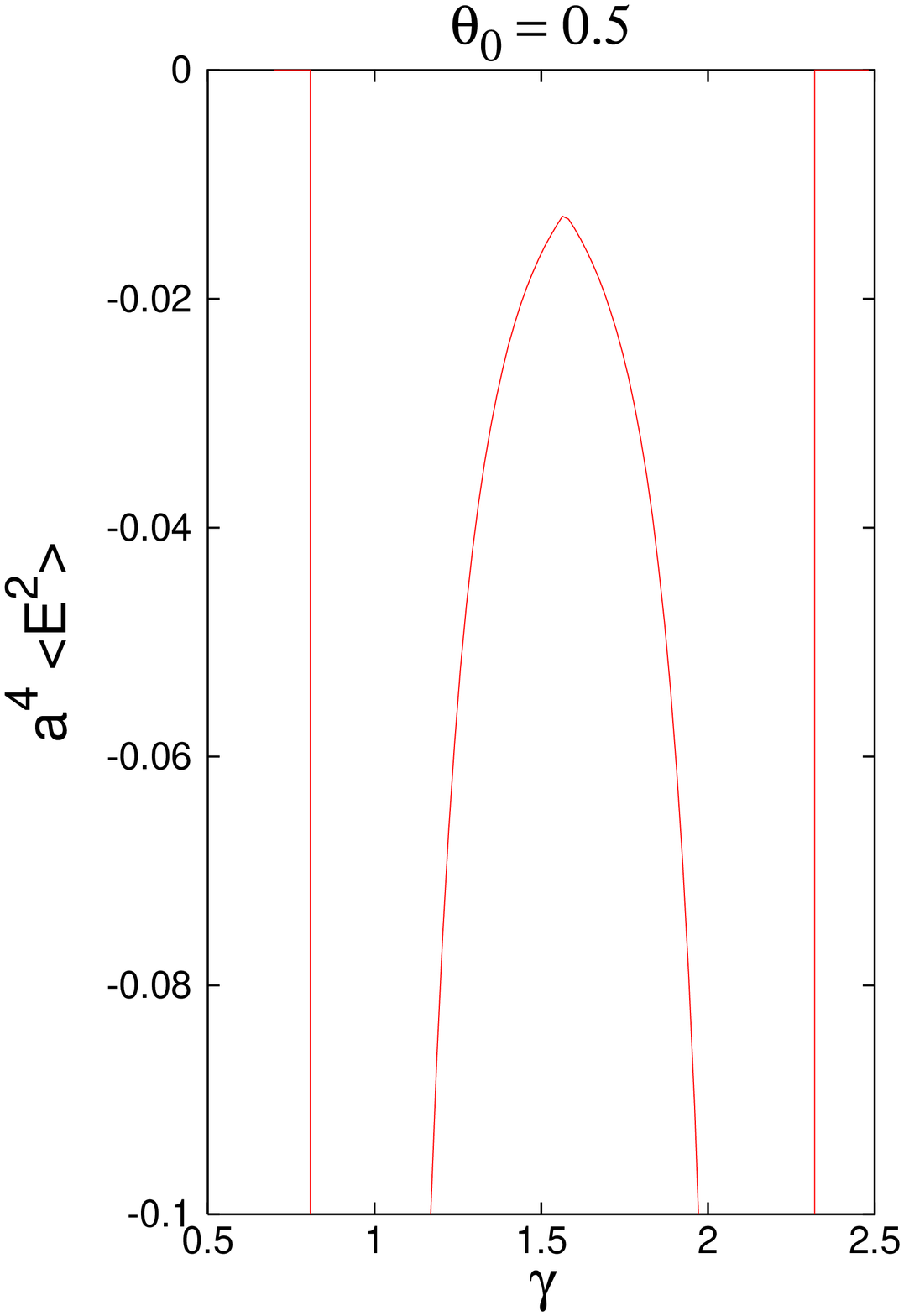 }
\end{center}
\caption{$\langle {\bf E}^2 \rangle$ for $\theta_0 = 0.5$. }
\label{fig:para_e_50}
\end{figure}

\begin{figure}
\begin{center}
\leavevmode\epsfysize=8cm\epsffile{ 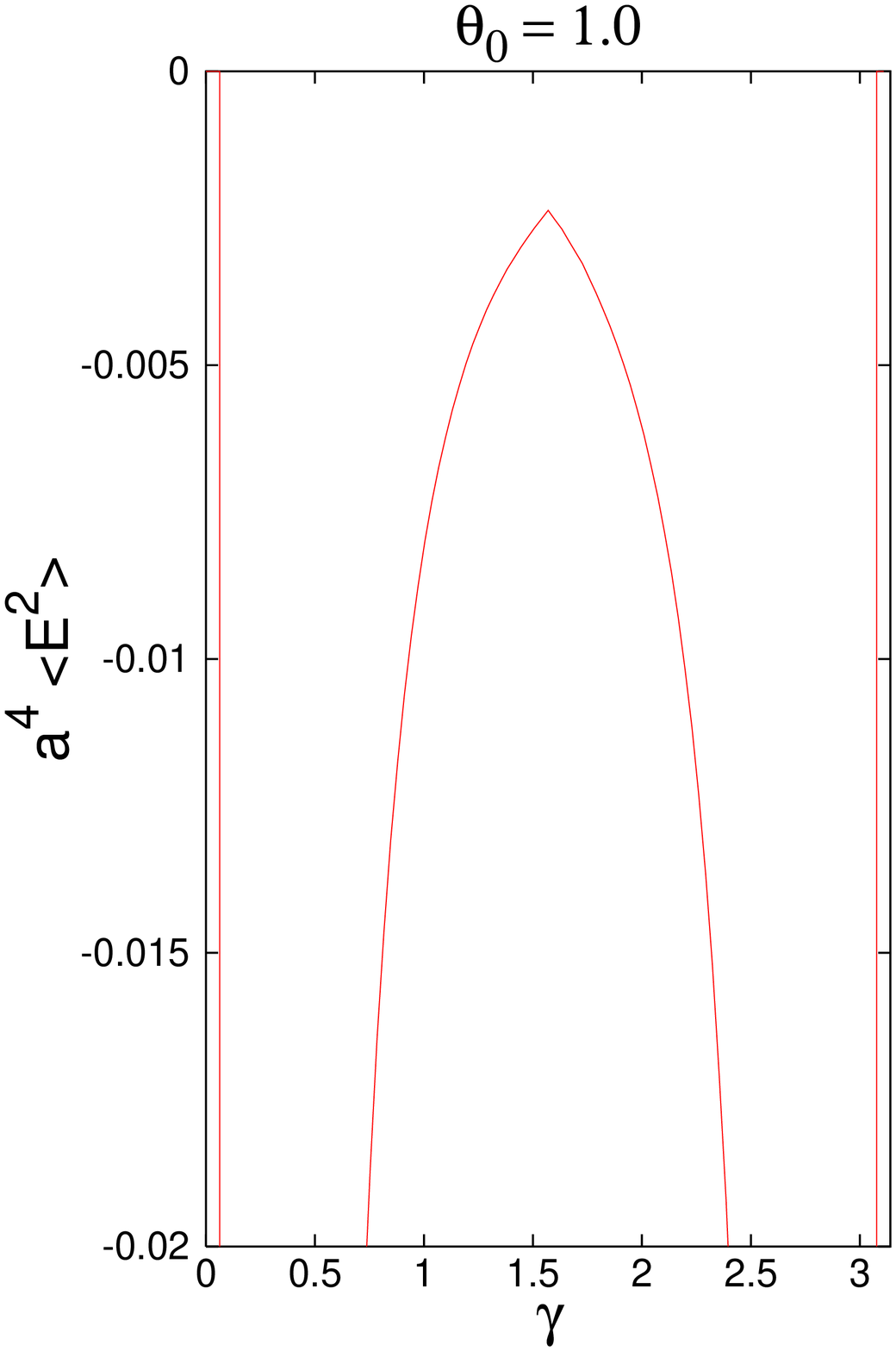}
\end{center}
\caption{$\langle {\bf E}^2 \rangle$ for $\theta_0 = 1.0$. }
\label{fig: para_e_100}
\end{figure}

\begin{figure}
\begin{center}
\leavevmode\epsfysize=8cm\epsffile{ 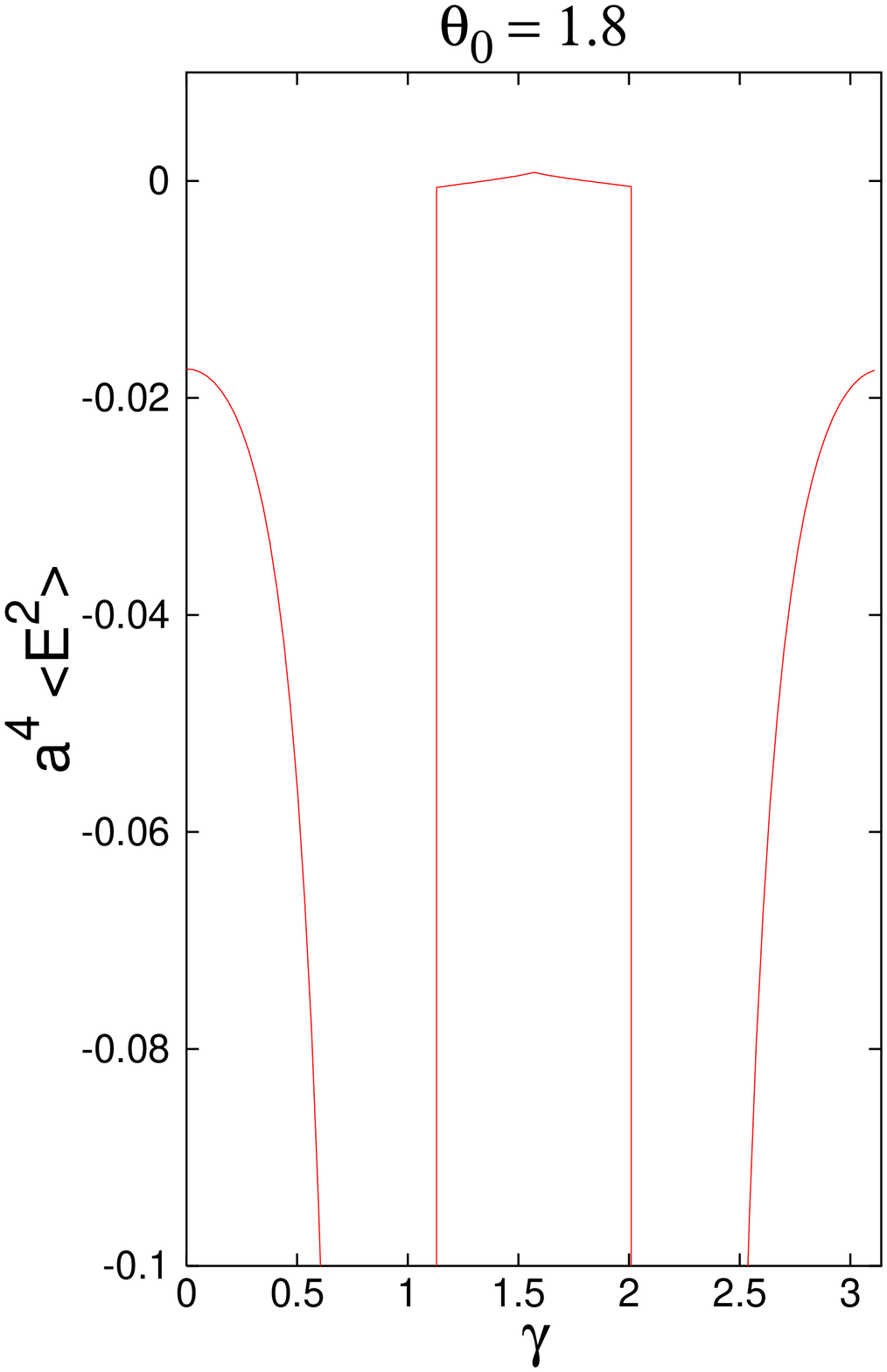}
\end{center}
\caption{$\langle {\bf E}^2 \rangle$ for $\theta_0 = 1.8$. }
\label{fig:para_e_180}
\end{figure}

\begin{figure}
\begin{center}
\leavevmode\epsfysize=8cm\epsffile{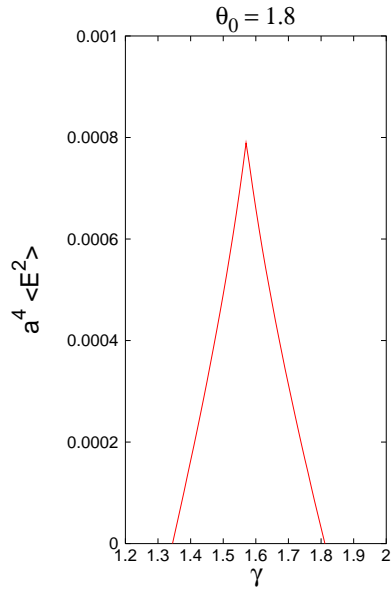 }
\end{center}
\caption{$\langle {\bf E}^2 \rangle$ for $\theta_0 = 1.8$, showing in detail
 the region where $\langle {\bf E}^2 \rangle > 0$. }
\label{fig:para_e_180_2}
\end{figure}

Numerical results for $\langle \varphi^2 \rangle$ and 
$\langle {\bf E}^2 \rangle$ for various values of $\theta_0$ are shown in 
Figs.~\ref{fig:para_p_50}-\ref{fig:para_e_180_2}
as functions of $\gamma$. In all cases, there are values of $\gamma$
at which $\langle \varphi^2 \rangle$ or $\langle {\bf E}^2 \rangle$
are singular in the model of a perfectly reflecting mirror with sharp edges.
 These points occur when 
an extremum of the function $f(\theta)$ sits at an edge of the mirror,
$\theta' = \pm \theta_0$. The values and slopes of $\langle \varphi^2 \rangle$ 
and $\langle {\bf E}^2 \rangle$ at $\gamma = \frac{\pi}{2}$ are in agreement
with Eqs.~(\ref{eq:phi_pi/2}) and (\ref{eq:E_pi/2}). We can see from the
graphs that for smaller mirrors, $\theta_0 < \frac{\pi}{2}$, we have
$\langle \varphi^2 \rangle > 0$ and $\langle {\bf E}^2 \rangle < 0$
everywhere. For larger mirrors, $\theta_0 > \frac{\pi}{2}$, we have regions 
where $\langle \varphi^2 \rangle >0$ and other regions where it is negative,
and similarly for $\langle {\bf E}^2 \rangle$. The signs of 
$\langle \varphi^2 \rangle$ and of $\langle {\bf E}^2 \rangle$ always
seem to be opposite. The regions near $\gamma = \frac{\pi}{2}$ for larger
mirrors where $\langle {\bf E}^2 \rangle > 0$ are of special interest.
These are regions where an atom will feel an attractive force toward the
focus, and thus has the possibility of being trapped. 

\section{Limits of Validity of the Results}
\label{sec:limits}

In this section, we will discuss the likely ranges of validity of the model
we have used to calculate $\langle \varphi^2 \rangle$ and
$\langle {\bf E}^2 \rangle$. In particular, we have assumed a geometric
optics approximation, and a mirror which is both perfectly reflecting
and has sharp edges. Each of these assumptions will be examined critically.

\subsection{Geometric Optics}

The use of geometric optics amounts to ignoring diffraction effects,
so we can gauge the accuracy of geometric optics by estimating the size
of these effects. In the geometric optics approximation, the incident
and reflected waves in Eq.~(\ref{eq:mode}) all have the same magnitude. 
Suppose that we now introduce a correction due to diffraction and write
\begin{equation}
F_{\bf k} = f_{\bf k} + \sum_i  f^{(i)}_{\bf k} + \Delta F_{\bf k}
         \, . \label{eq:mode2}
\end{equation}
In the case of a plane strip which has width $2 y_0$ in one direction and 
is infinite in the other direction, the magnitude of the diffraction
correction is estimated in the Appendix with the result 
$|\Delta F_{\bf k}/F_{\bf k}| \approx \sqrt{\lambda/y_0}$, where $\lambda$
is the wavelength of the incident wave. In our case, we have a parabolic
cylinder characterized by the length scale $b$. We could imagine 
representing the  parabolic cylinder by a set of strips which are infinite 
in the $z$-direction. The order of magnitude of the diffraction correction
should be the same as for a single strip of width $b$. Thus we estimate
that in the present case
\begin{equation}
\left|\frac{\Delta F_{\bf k}}{F_{\bf k}}\right| \approx 
              \sqrt{\frac{\lambda}{b}}  \,.  \label{eq:diff_ratio}
\end{equation}

The wavelengths which give the dominant contribution to 
$\langle \varphi^2 \rangle$ and $\langle {\bf E}^2 \rangle$ near the focus
are those of order $a$. Thus the interference term between $\Delta F_{\bf k}$
and any of the other terms in Eq.~(\ref{eq:mode2}) should yield contributions
to $\langle \varphi^2 \rangle$ and $\langle {\bf E}^2 \rangle$ which are
smaller than the dominant contribution by a factor of the order of
$\sqrt{a/b}$. Thus we estimate the diffraction contribution to
$\langle \varphi^2 \rangle$ to be of order 
 \begin{equation}
\Delta \langle \varphi^2 \rangle \approx 
       \frac{1}{a^{\frac{3}{2}}\,b^{\frac{1}{2}}} \, ,   \label{eq:diff_phi}
\end{equation}
and that to $\langle {\bf E}^2 \rangle$ to be of order
\begin{equation}
\Delta \langle {\bf E}^2 \rangle \approx 
           \frac{1}{a^{\frac{7}{2}}\,b^{\frac{1}{2}}} \, .  \label{eq:diff_E}
\end{equation}
Note that the geometric optics results near the focus depend only upon
$a$ and $\theta_0$, the angular size of the mirror, but not upon $b$, the linear
dimension of the mirror. However, the diffraction correction decreases
with increasing $b$ and hence can be made smaller for a larger, more
distant mirror.

In classical optics, the effects of diffraction are normally of the order
of the wavelength divided by the size of the object. This arises when one
is looking at the power in the diffracted wave, which is given by the
square of Eq.~(\ref{eq:diff_ratio}). In our case, there is a possible 
contribution from an interference term between the diffracted wave and 
the geometric optics contributions. If a more detailed 
calculation were to find that this
term vanished, then our estimates for $\Delta \langle \varphi^2 \rangle$
and $\Delta \langle {\bf E}^2 \rangle$ would become smaller than
Eqs.~(\ref{eq:diff_phi}) and (\ref{eq:diff_E}), respectively, by an 
additional factor of $\sqrt{a/b}$.

\subsection{Edge Effects and Finite Reflectivity}

Our model of the mirror is one in which it is not only perfectly reflecting, but
also has a sharp edge at which the reflectivity falls discontinuously to
zero. Both of these assumption are over simplifications to which the
singular behavior of  $\langle \varphi^2 \rangle$ and 
$\langle {\bf E}^2 \rangle$  may be attributed. Consider the sharp edge
assumption. A more realistic model might have the reflectivity falling
smoothly to zero over an angular interval of width $\Delta \theta$ at the
edges of the mirror. This would remove the singularities found above 
at specific values of $\gamma$. Recall that these singularities arise when
an edge of the mirror, at $\pm \theta_0$, sits at an extremum of $f(\theta')$.
In the case of a smoothed edge, both $\langle \varphi^2 \rangle$ and 
$\langle {\bf E}^2 \rangle$ will be bounded for fixed $a$
\begin{equation}
|\langle \varphi^2 \rangle| < \frac{1}{a^2\, \Delta \theta}
\end{equation}
and 
\begin{equation}
|\langle {\bf E}^2 \rangle| < \frac{1}{a^4\, (\Delta \theta)^3}  \, .
\end{equation}
Smoothing of the edges of the mirror is also expected to remove the
cusps at $\gamma = \frac{\pi}{2}$.

The smoothed edges do not, however, remove the singularities as
$a \rightarrow 0$. This singularity is presumably due to the assumption
of perfect reflectivity at all wavelengths. A more realistic model would
have the reflectivity go to zero at short  wavelengths. Suppose that the
mirror becomes transparent for wavelengths less than some minimum value,
$\lambda_m$. Then we expect to find the bounds
\begin{equation}
|\langle \varphi^2 \rangle| < \frac{1}{\lambda_m^2}  \label{eq:phi_bound}
\end{equation}
and 
\begin{equation}
|\langle {\bf E}^2 \rangle| < \frac{1}{\lambda_m^4}  \, . \label{eq:E_bound}
\end{equation}
One reason for reduced reflectivity at short  wavelengths is dispersion.
The mirror can be regarded as close to perfectly reflecting only for
wavelengths longer than about the plasma wavelength $\lambda_P$ of the 
metal in the mirror. Thus our results are only valid when $a > \lambda_P$.
For aluminum, for example, $\lambda_P \approx 84 {\rm nm}$. However,
it is unlikely that dispersion alone is capable of removing all short
distance singularities. The reason for this is that dielectric functions
approach unity as $(\lambda/\lambda_P)^2$ as $\lambda \rightarrow 0$. 
This is not fast enough to regulate integrals which diverge quartically
as the short  wavelength cutoff. In the case of a plane interface between
vacuum and a dispersive material, quantities such as $\langle {\bf E}^2 \rangle$
still diverge as the interface is approached, although less rapidly than
in the case of a perfect mirror \cite{SF}. Thus it seem that other effects, 
such as surface roughness, the breakdown of the continuum description
at the atomic level, or quantum uncertainty in the location of the 
interface \cite{FS98} are needed to produce finite values of 
$\langle {\bf E}^2 \rangle$. If any effect does produce a sufficiently
sharp cutoff at short wavelengths below about $\lambda_m$, then we would 
have the bounds in Eqs.~(\ref{eq:phi_bound}) and (\ref{eq:E_bound})
even without removing the assumption of a sharp edge to the mirror.

\section{Observable Consequences}
\label{sec:obs}

In I, several possible experimental tests of the enhanced vacuum fluctuation
near the focus were discussed. Here these tests will be reviewed and further
discussed. The most direct test would seem to be to measure the force on
an atom or other polarizable particle. In a regime where the atom can be 
described by a static polarizability $\alpha$, the force is 
${\bf F} = - {\bf \nabla} V$, where
\begin{equation}
V = - \frac{1}{2}\, \alpha\, \langle {\bf E}^2 \rangle \,. \label{eq:CP}
\end{equation}
In the vicinity of the focus, we have found that
\begin{equation}
\langle {\bf E}^2 \rangle = \frac{\Lambda}{a^4} \, ,
\end{equation}
where $\Lambda$ is a dimensionless constant. For an atom in its ground state,
Eq.~(\ref{eq:CP}) should be a good approximation when $a$ is larger than the
wavelength associated with the transition to the first excited state. 

One might try to measure the defection of atoms moving parallel to the focal
line of a parabolic cylinder. The analogous experiment for a flat plate
was performed by Sukenik {\it et al} \cite{Sukenik} and confirmed  Casimir
and Polder's theoretical prediction \cite{CP}. In the present case, the expected
angular deflection is
\begin{equation}
\frac{\Delta a}{a} = 0.25\, \left(\frac{\Lambda}{10^{-3}} \right)\, 
\left(\frac{\alpha}{\alpha_{Na}} \right)\,
 \left(\frac{m_{Na}}{m} \right)\, \left(\frac{1 \mu {\rm m}}{a} \right)^6\, 
\left(\frac{t}{10^{-3} s} \right)^2 \,.
\end{equation}
Here $m_{Na}=3.8 \times 10^{-23} {\rm gm}$ and 
$\alpha_{Na}=3.0\times 10^{-22} {\rm cm}^3$ denote the mass and polarizability 
of the sodium atom, respectively. (Note that polarizability in the 
Lorentz-Heaviside which we use is $4 \pi$ times that in Gaussian units.) 
If $t$ is of order $10^{-3} s$ (the time needed
for an atom with a kinetic energy of order 300K to travel a few centimeters),
and $z$ is of order $1 \mu {\rm m}$, the fractional deflection is significant.

An alternative to measuring the deflection of the beam might be to measure
the relative phase shift in an atom interferometer, which are sensitive to
phase shifts of the order of $10^{-4}$ radians \cite{WPW}. If one path of
the interferometer were to be parallel to the focal line at a mean distance
of $a$ and the other path far away, the phase difference will be
\begin{equation}
\Delta \phi = \frac{t}{2}\, \alpha\, \langle {\bf E}^2 \rangle 
= 0.04 \,\left(\frac{\Lambda}{10^{-3}} \right)\,
\left(\frac{\alpha}{\alpha_{Na}} \right)\,
\left(\frac{1 \mu {\rm m}}{a} \right)^4 \,\left(\frac{t}{10^{-3} s} \right) \, .
\end{equation}
Both deflection and phase shift measurements would require the atoms to
be rather well collimated and localized near the focal line.

Perhaps the most dramatic confirmation of enhanced vacuum fluctuations would be
the trapping of atoms near the focus. This would require a mirror with
$\theta_0 > \frac{\pi}{2}$ so that there is a region with 
$\langle {\bf E}^2 \rangle > 0$. It would also require the atoms to be
cooled below a temperature of about
\begin{equation}
T_m = 2 \times 10^{-9} K\, \left(\frac{\Lambda}{10^{-3}} \right)\, 
\left(\frac{\alpha}{\alpha_{Na}} \right)\,
\left(\frac{1 \mu {\rm m}}{a} \right)^4 \,.
\end{equation}
Thus at temperatures of the order of $10^{-9} K$, atoms could become trapped
in a region of the order of $1 \mu {\rm m}$ from the focus. This type
of trapping would be quite different from that currently employed
\cite{Phillips} in that it would require no applied classical electromagnetic
fields. 

\begin{figure}
\begin{center}
\leavevmode\epsfysize=6cm\epsffile{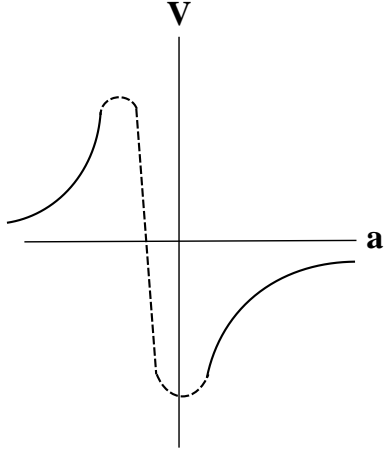}
\end{center}
\caption{A possible form of the potential energy function $V(a,\gamma)$ is
illustrated. The solid lines show the region where $V$ can be reliably
computed within our approximations. 
The solid line on the right is the form of $V$ as a function 
of $a$ for a value of $\gamma$ for which the force is attractive. The
solid line on the left is for a value (not necessarily in the opposite
direction) for which it is repulsive. The dashed line is a possible form
which interpolates between these two regions, but which would have to be
computed by a more detailed theory. }
\label{fig:potential}
\end{figure}

The peculiar property that the force on an atom can be attractive from certain
directions and repulsive from other directions requires some discussion.
If one approaches from an attractive direction, the potential energy is 
becoming increasingly negative, $V \sim -1/a^4$, whereas from a repulsive
direction it is becoming large and positive, $V \sim +1/a^4$. In both
cases, $V \rightarrow 0$ for large $a$. At some minimum value of $a$,
the inverse fourth power behavior has to be modified. Thus the exact
potential energy $V(a,\gamma)$ must be a continuous function with a
stable minimum, as illustrated in Fig.~\ref{fig:potential}. Note that the
requirement that $V \rightarrow 0$ for large $a$ in all directions rules
out the possibility that there is only a saddle point, and no minimum.
Although we can infer the existence
of the minimum from our results, the detailed form of $V(a,\gamma)$ near
the minimum could only be found by an analysis which goes beyond the
approximations made in the present paper. 

\section{Discussion and Conclusions}
\label{sec:final}

In this paper we have further developed a geometric optics approach to the 
study of vacuum fluctuations near the focus of parabolic mirrors. The main
result of this approach is that the mean squared scalar field 
$\langle \varphi^2 \rangle$ and mean squared electric field
$\langle {\bf E}^2 \rangle$ grow as inverse powers of $a$, the distance from the
focus, for small $a$. The key justification of geometric optics is its
self-consistency. When $\langle \varphi^2 \rangle$ and 
$\langle {\bf E}^2 \rangle$ are large, the dominant contributions must come 
from short wavelengths for which geometric optics is a good approximation.
This was discussed more quantitatively in Sect.~\ref{sec:limits}. 

We have given some explicit analytic and numerical results for 
$\langle \varphi^2 \rangle$ and $\langle {\bf E}^2 \rangle$ near the
focal line of a parabolic cylindrical mirror. In this paper, we restricted
our attention to the case that the angular size of the mirror, $2 \theta_0$,
is less than $\frac{4 \pi}{3}$. This insures that there are never more than
two reflected rays for a given incident ray and simplifies the analysis 
We find that for smaller mirrors, $\theta_0 < \frac{\pi}{2}$, 
$\langle \varphi^2 \rangle > 0$ and $\langle {\bf E}^2 \rangle <0$ everywhere.
In this case, an atom will feel a repulsive force away from the focus.
For larger mirrors, $\frac{\pi}{2} < \theta_0 < \frac{2 \pi}{3}$, these
mean squared quantities can have either sign, depending upon the direction from
the focus. In directions nearly perpendicular to the symmetry axis,
$\gamma \approx \frac{\pi}{2}$, we find $\langle \varphi^2 \rangle < 0$ and 
$\langle {\bf E}^2 \rangle > 0$. In this case, the force on an atom is 
attractive toward the focus and trapping becomes a possibility.

In the geometric optics approximation, the mean squared electric and
magnetic fields are equal, so the local energy density is equal to
$\langle {\bf E}^2 \rangle$. Thus, when $\langle {\bf E}^2 \rangle < 0$, 
the local energy density is negative, and one has columns of negative
energy density running parallel to the focal line of the parabolic cylinder.

The feasibility of experiments to observe the effects on atoms near the focus 
was discussed in Sect.~\ref{sec:obs}. Although the effects are small,
it seems plausible that they could be observed.

\begin{acknowledgments}
  We would like to thank Ken Olum for valuable 
discussions. This work was supported in part by the National
Science Foundation under Grant PHY-9800965, 
by Conselho Nacional de Desenvolvimento
Cientifico e Tecnologico do Brasil (CNPq), and by 
the U.S.Department of Energy (D.O.E.) under cooperative research agreement
DF-FC02-94ER40810.
\end{acknowledgments}

\appendix*
\section{}

In this Appendix, we will develop a method which goes beyond the geometric 
optics approximation, and use it to estimate the size of the corrections
due to diffraction effects. Here we will discuss only a scalar field which
satisfies Dirichlet boundary conditions on the mirror. The basic method
is to write down and then approximately solve an integral equation for the 
scattered  wave. This method was first used by Kirchhoff to discuss
diffraction in classical optics. Let $\phi({\bf x})$ be a solution of the
Helmholtz equation
\begin{equation}
\nabla^2 \phi + k^2\, \phi = 0 \, , \label{eq:phieq}
\end{equation}
and let $G({\bf x},{\bf x}')$ be a Green's function for this equation
\begin{equation}
\nabla^2 G + k^2\, G = \delta({\bf x}-{\bf x}') \, . \label{eq:Geq}
\end{equation}
If we multiply Eq.~(\ref{eq:phieq}) by $G$, Eq.~(\ref{eq:Geq}) by $\phi$,
take the difference, and integrate over an arbitrary spatial volume $V$,
the result is
\begin{equation}
\phi({\bf x}') = \int_S d{\bf a}\cdot 
[\phi({\bf x})\,{\bf \nabla}G({\bf x},{\bf x}') -   
G({\bf x},{\bf x}')\, {\bf \nabla} \phi({\bf x})] \,.
\end{equation}
Here $S$ is the boundary of $V$, and $d{\bf a}$ is outward directed.

\begin{figure}
\begin{center}
\leavevmode\epsfysize=8cm\epsffile{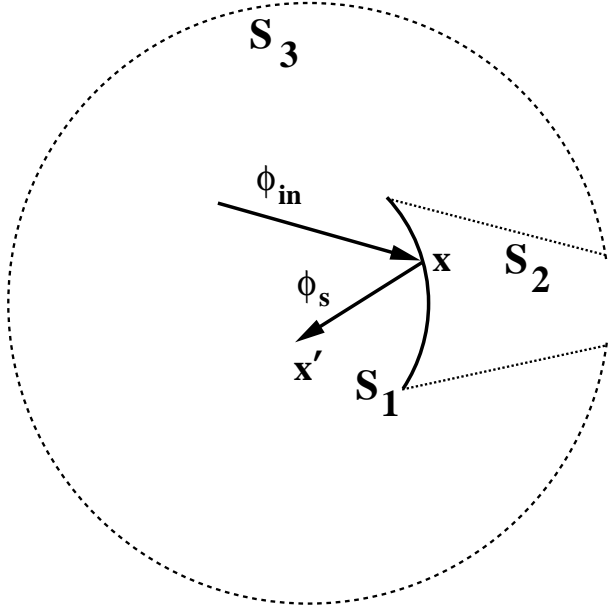}
\end{center}
\caption{A closed surface $S$ consists of three parts, the open surface
of interest $S_1$, a segment $S_2$ hidden from the view of an observer at
${\bf x}'$, and a portion $S_3$ at a very large distance. An incident wave
$\phi_{in}$ reflects from  $S_1$ and produces the scattered wave $\phi_{s}$. 
Here ${\bf x}$ is an arbitrary point in $S_1$. }
\label{fig:surface}
\end{figure}
We are interested in calculating the scattered wave from a open surface
$S_1$, rather than the solution on the interior of a closed surface $S$.
However, we can let the closed surface $S$ consist of three segments,
as illustrated in Fig.~\ref{fig:surface}. 
The first is the surface of interest $S_1$, the
second is a segment $S_2$ which is hidden from the view of an observer
in the region where we wish to find the scattered wave, and the third
$S_3$ closes the surface at a large distance. We now ignore the contributions 
of $S_2$ and $S_3$, and assume that $\phi=0$ on $S_1$, so that we have
\begin{equation}
\phi({\bf x}') = - \int_{S_1} d{\bf a}\cdot {\bf \nabla} \phi({\bf x})
                      \, G({\bf x},{\bf x}') \, .  \label{eq:inteq}
\end{equation}
This is still an integral equation which relates values of $\phi$
on the surface to those off of the surface. We can solve it in an approximation 
in which an incident wave scatters only once from the surface. Let
$\phi_{in}$ be the incident wave, and assume that on $S_1$
\begin{equation}
{\bf \hat{n}}\cdot {\bf \nabla} \phi = 
2 {\bf \hat{n}}\cdot {\bf \nabla} \phi_{in} \, .
\end{equation} 
We next choose the empty space Green's function
\begin{equation}
G({\bf x},{\bf x}') = 
- \frac{{\rm e}^{i k |{\bf x}-{\bf x}'|}}{4 \pi |{\bf x}-{\bf x}'|} \,,
\end{equation}
and interpret Eq.~(\ref{eq:inteq}) as giving the scattered wave, $\phi_{s}$,
\begin{equation}
\phi_s({\bf x}') = 
   - 2\int_{S_1} d{\bf a}\cdot {\bf \nabla} \phi_{in}({\bf x})
                      \, G({\bf x},{\bf x}') \, .  \label{eq:scat}
\end{equation}
The physical interpretation of this expression is that each point on
$S_1$ radiates a scattered wave proportional to  
${\bf \hat{n}}\cdot {\bf \nabla} \phi_{in}$; the superposition of these
individual contributions forms the net scattered wave, as expected from
Huygen's principle. 

\begin{figure}
\begin{center}
\leavevmode\epsfysize=8cm\epsffile{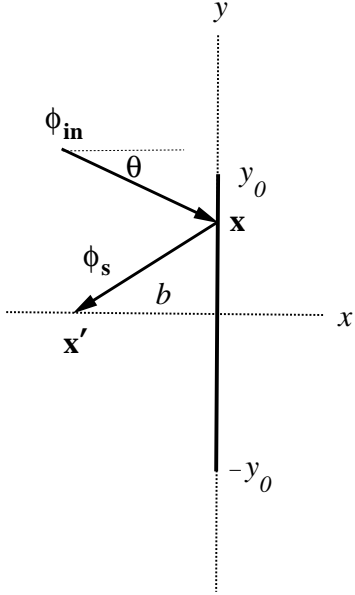}
\end{center}
\caption{A reflecting strip lies in the interval $-y_0 < y < y_0$ and
is infinite in the $z$-direction, normal to the page. A plane wave
$\phi_{in}$ is incident at an angle of $\theta$ and creates the reflected
wave $\phi_{s}$ at the point ${\bf x}'$, located a distance $b$ from the strip. }
\label{fig:strip}
\end{figure}

Let the incident wave be a plane wave
\begin{equation}
\phi_{in} = {\rm e}^{i {\bf k}\cdot {\bf x}} \, ,
\end{equation}
and the surface $S_1$ be the portion of the $x=0$ plane in the interval
$-y_0 < y < y_0$. That is, it is a strip which is infinite in the $z$-direction.
Take the wavevector of the incident wave to be 
${\bf k} = k(\cos\theta,-\sin\theta,0)$ and the observation point to be
${\bf x}' = (-b,0,0)$, that is, at a distance $b$ from the mirror, as
illustrated in Fig.~\ref{fig:strip}. We can
now write the scattered wave as
\begin{equation}
\phi_s({\bf x}') =  \frac{i k \cos\theta}{2 \pi} \int_{-y_0}^{y_0} d y\,
{\rm e}^{-i k \sin\theta \,y}  \int_{-\infty}^{\infty} dz \,
\frac{{\rm e}^{i k \sqrt{y^2+z^2+b^2}}}{\sqrt{y^2+z^2+b^2}} \,.
\end{equation}
The $z$-integration may be performed explicitly, with the result
\begin{equation}
\phi_s({\bf x}') = -\frac{1}{2} k \cos\theta \int_{-y_0}^{y_0} d y\,
{\rm e}^{-i k \sin\theta \,y} \, H_0^{(1)}(k\sqrt{y^2+b^2}) \, ,
                                             \label{eq:scat2}
\end{equation}
where $H_0^{(1)}$ is the Hankel function of the first kind. In general, 
the remaining integral cannot be evaluated explicitly. However, in the
case of an infinite plane mirror, $y_0 \rightarrow \infty$, it can be
evaluated, with the result
\begin{equation}
\phi_s({\bf x}') = - {\rm e}^{i k \sin\theta b}  
= - {\rm e}^{i {\bf k}'\cdot {\bf x}'} \,. \label{eq:scat3}
\end{equation}
This is just the result predicted by geometric optics. In the special case
of an infinite plane mirror, geometric optics gives the exact result. 
In the case of a finite mirror, Eq.~(\ref{eq:scat3}) is still a good 
approximation in the high frequency limit. In this limit, one may evaluate
the integral in Eq.~(\ref{eq:scat2}) using the stationary phase approximation.
So long as the classical path intersects the mirror in the interval
$-y_0 < y < y_0$, then there is one point of stationary phase within the
range of integration. This intersection occurs if $-y_0 < b \cos\theta < y_0$.
The result of the stationary phase approximation is just Eq.~(\ref{eq:scat3}),
reflecting the fact that geometric optics is a good approximation in the
high frequency limit. 

Now we wish to give a quantitative estimate of the accuracy of the 
approximation. Let
\begin{equation}
\Delta \phi_s = \frac{1}{2} k \cos\theta \left(\int_{-\infty}^{-y_0}
+ \int_{y_0}^{\infty}\right)d y\,
{\rm e}^{-i k \sin\theta \,y} \, H_0^{(1)}(k\sqrt{y^2+b^2}) \, .
                                             \label{eq:scat4}
\end{equation}
This is the difference between the stationary phase (geometric optics)
approximation and the exact result for a finite mirror, so 
$|\Delta \phi_s| = |\Delta \phi_s|/|\phi_s|$ is a fractional measure of 
the accuracy of the approximation.

In the high frequency limit that $k \gg b$, we can use the large argument
form for $H_0^{(1)}$,
\begin{equation}
H_0^{(1)}(z) \approx \sqrt{\frac{2}{\pi z}} \, {\rm e}^{i(z-\frac{\pi}{4})} \, ,
\end{equation}
and write
\begin{equation} 
\Delta \phi_s \approx \sqrt{\frac{k}{2 \pi}}\,\cos\theta \,{\rm e}^{-i \pi/4}
\int_{y_0}^{\infty} d y\, (y^2+b^2)^{-\frac{1}{4}}\,
{\rm e}^{-i k (\sin\theta \,y - \sqrt{y^2+b^2})} \; + (y_0 \rightarrow -y_0)\,.
\end{equation}
The above integral still cannot be evaluated explicitly, but we can
estimate it as being of order $1/(k \sqrt{y_0})$ when $k \ll y_0$. With
this estimate, we find that 
\begin{equation} 
|\Delta \phi_s| \sim O\left(\frac{\lambda}{y_0}\right)^{\frac{1}{2}} \, ,
\end{equation}
where $\lambda$ is the wavelength of the incident wave.

\end{document}